\documentclass[12pt]{emulateapj}

\usepackage{amsmath,amssymb}  
\usepackage{graphicx}

\newcommand{\eg}{{\it e.g.}}

\slugcomment{ }
 
\shorttitle{The integrated stellar content of dark matter halos}
\shortauthors{Leauthaud et al.}

\begin{document}
  

 \title{The integrated stellar content of dark matter halos}




 \author{Alexie Leauthaud\altaffilmark{1,2}, Matthew
   R. George\altaffilmark{3}, Peter S. Behroozi\altaffilmark{4}, Kevin
   Bundy\altaffilmark{3}, Jeremy Tinker\altaffilmark{5}, Risa
   H. Wechsler\altaffilmark{4}, Charlie Conroy\altaffilmark{6}, Alexis
   Finoguenov\altaffilmark{7,8}, Masayuki Tanaka\altaffilmark{9}}

\submitted{Submitted to ApJ}
\email{asleauthaud@lbl.gov}

\altaffiltext{1}{Lawrence Berkeley National Lab, 1 Cyclotron Road,
  Berkeley, CA 94720, USA}

\altaffiltext{2}{Berkeley Center for Cosmological Physics, University
  of California, Berkeley, CA 94720, USA}

\altaffiltext{3}{Department of Astronomy, University of California,
  Berkeley, CA 94720, USA}

\altaffiltext{4}{Kavli Institute for Particle Astrophysics and
  Cosmology; Physics Department, Stanford University, and SLAC
  National Accelerator Laboratory, Stanford, CA 94305}

\altaffiltext{5}{Center for Cosmology and Particle Physics, Department
  of Physics, New York University}

\altaffiltext{6}{Harvard-Smithsonian Center for Astrophysics, Cambridge, MA,USA}

\altaffiltext{7}{Max Planck Institut f\"{u}r extraterrestrische
  Physik, Giessenbachstrasse, D-85748 Garchingbei M\"{u}nchen,
  Germany}

\altaffiltext{8}{University of Maryland Baltimore County, 1000 Hilltop
  circle, Baltimore, MD 21250, USA}

\altaffiltext{9}{Institute for the Physics and Mathematics of the Universe, The University of Tokyo, 5-1-5 Kashiwanoha, Kashiwa-shi, Chiba 277-8583, Japan}

  
\begin{abstract} Measurements of the total amount of stars locked up
  in galaxies as a function of host halo mass contain key clues about
  the efficiency of processes that regulate star formation.  We derive
  the total stellar mass fraction $f_{\star}$ as a function of halo
  mass $M_{500c}$ from $z=0.2$ to $z=1$ using two complementary
  methods. First, we derive $f_{\star}$ using a statistical Halo
  Occupation Distribution model jointly constrained by data from
  lensing, clustering, and the stellar mass function. This method
  enables us to probe $f_{\star}$ over a much wider halo mass range
  than with group or cluster catalogs. Second, we derive $f_{\star}$
  at group scales using a COSMOS X-ray group catalog and we show that
  the two methods agree to within 30\%. We quantify the systematic
  uncertainty on $f_{\star}$ using abundance matching methods and we
  show that the statistical uncertainty on $f_{\star}$ ($\sim 10$\%)
  is dwarfed by systematic uncertainties associated with stellar mass
  measurements ($\sim 45$\% excluding IMF uncertainties). Assuming a
  Chabrier IMF, we find $0.012\leq f_{\star} \leq 0.025$ at
  $M_{500c}=10^{13}$ M$_{\odot}$ and $0.0057\leq f_{\star} \leq 0.015$
  at $M_{500c}=10^{14}$ M$_{\odot}$. These values are significantly
  lower than previously published estimates. We investigate the cause
  of this difference and find that previous work has overestimated
  $f_{\star}$ due to a combination of inaccurate stellar mass
  estimators and/or because they have assumed that all galaxies in
  groups are early type galaxies with a constant mass-to-light
  ratio. Contrary to previous claims, our results suggest that the
  mean value of $f_{\star}$ is always significantly lower than $f_{\rm
    gas}$ for halos above $10^{13} M_{\odot}$. Combining our results
  with recently published gas mas fractions, we find a shortfall in
  $f_{\star}$+$f_{\rm gas}$ at $R_{500c}$ compared to the cosmic
  mean. This shortfall varies with halo mass and becomes larger
  towards lower halos masses. \end{abstract}
 

 

\section{Introduction}

Dark matter halos contain baryons in the forms of stars and gas, but
there is active debate about the level of agreement between the baryon
fraction in halos and the cosmic mean, $f_b \equiv
\frac{\Omega_b}{\Omega_m}$, as well as which baryonic phase dominates
as a function of halo mass $M_h$. Cosmological simulations suggest
that the baryon content of massive galaxy clusters (M$_h \sim 10^{15}$
M$_{\odot}$) should be within 10\% of the universal value
\citep[\eg,][]{Kravtsov:2005,Ettori:2006}. In galaxy clusters, most
baryons reside in the hot, diffuse, X-ray emitting intra-cluster
medium (ICM). Detailed X-ray measurements suggest that the hot gas
fraction ($f_{\rm gas}$) is considerably lower than the cosmic value
in the inner regions of clusters
\citep[\eg,][]{Ettori:2003,Lin:2003,Vikhlinin:2006,McCarthy:2007,Arnaud:2007,Allen:2008,Sun:2009}
though recent measurements suggest that $f_{\rm gas}$ might approach
the universal value at larger radii \citep[][]{Simionescu:2011}.

Depleted gas mass fractions at small radii might indicate that the ICM
has been affected by non gravitational processes such as star
formation or feedback from Active Galatic Nuclei (AGN). Indeed,
sources of heat such as these that are located near the centers of
clusters might pump thermal energy into the ICM causing it to expand
towards the outskirts (thus lowering $f_{\rm gas}$ in cluster
cores). Significant sources of heat might even be capable of removing
gas from lower-mass halos \citep[][]{McCarthy:2011}. On the other
hand, the fact that $f_{\rm gas}$ is lower than expected at small
radii might also simply indicate that some gas has been transformed
into other baryonic components such as galaxy stellar mass or
Intra-Cluster Light (ICL). Discriminating between these two scenarios
is key towards understanding the efficiency of processes that regulate
star formation and for tracing the thermo-dynamic history of ICL.

Galaxy groups are highly interesting in this respect since they have
shallower potential wells than clusters and should therefore be more
sensitive to non-gravitational processes
\citep[\eg,][]{McCarthy:2011}. Measurements of the baryon fraction at
group scales, however, are disparate and subject to much debate
\citep[\eg,][]{Lin:2003, Gonzalez:2007,Giodini:2009,Balogh:2007}. In
particular, one unresolved issue is whether or not the baryon content
of galaxy groups is dominated by hot gas (like clusters) or if other
baryonic phases such as galaxy stellar mass or ICL also make a
significant contribution. From X-ray studies there is a fairly clear
consensus that $f_{\rm gas}$ decreases towards lower halo masses
\citep[\eg,][]{Vikhlinin:2006,Arnaud:2007,Sun:2009, Pratt:2009}. The
two other components, $f_{\star}$ and $f_{\rm ICL}$, have proved more
challenging to measure and are largely responsible for disagreements
in the literature regarding the group scale baryon fraction. The ICL
component is diffuse and faint (with a typical surface brightness of
27-32 mag arcsec$^{-2}$ in the r-band) and therefore extremely
difficult to measure
\citep[\eg,][]{Feldmeier:2004,Lin:2004,Zibetti:2005,Gonzalez:2005,Krick:2006}. The
$f_{\star}$ component is subject to uncertainties regarding the
assignment of galaxies to groups and clusters, but also to
uncertainties inherent to stellar mass measurements themselves.

Previous results on this topic can be broadly divided into two
categories. In the first category, the decrease in $f_{\rm gas}$
towards lower halo masses is found to be exactly compensated by an
increased contribution from $f_{\star}$+$f_{\rm ICL}$ in such a way
that the baryon fraction is constant with halo mass and is found to be
either slightly below \citep[][]{Gonzalez:2007} or significantly below
$f_b$ \citep[][]{Andreon:2010}. This would imply that there is a
simple trade-off between $f_{\rm gas}$ and $f_{\star}$+$f_{\rm ICL}$
from group to cluster scales.  In the second category,
$f_{\star}$+$f_{\rm ICL}$ do not exactly compensate for the decrease
of $f_{\rm gas}$ at group scales and thus the baryon fraction is found
to be mildly decreasing towards lower halo masses
\citep[\eg,][]{Lin:2003,Giodini:2009}. This could imply for example
that the ICM is more strongly affected by feedback mechanisms in
groups than in clusters.

Most of the work to date has not directly probed $f_{\star}$, but has
focused instead on the relationship between halo mass and the total
K-band luminosity of group and cluster galaxies
\citep[\eg,][]{Lin:2003,Ramella:2004,Balogh:2007,Balogh:2010}.
Although K-band luminosity is most sensitive to the low-mass stars
that dominate the total mass of most stellar populations, it is not a
direct tracer of this mass. Other work, for lack of multiband
photometry, has estimated $f_{\star}$ using simple mass-to-light ratio
(M/L) estimates \citep[\eg][]{Lin:2003,
  Gonzalez:2007,Balogh:2007,Lagana:2008,Giodini:2009,Andreon:2010,Balogh:2010,Dai:2010,Lagana:2011}. In
practice, however, mass-to-light ratio values can vary strongly with
galaxy color, metallicity, and age. No single mass-to-light ratio can
capture the complexity of the full galaxy population
\citep[][]{Ilbert:2009}. No studies to date have actually accounted
for the full morphological mix of galaxies in group environments and
used Spectral Energy Distribution (SED) fitting methods to derived
$f_{\star}$.

In this paper we measure $f_{\star}$ using stellar masses derived from
full SED fitting to multi-band photometry (including K-band) from the
COSMOS survey. We tackle the issue of measuring $f_{\star}$ using two
complementary approaches. First, we use a Halo Occupation Distribution
(HOD) model that has been calibrated to fit measurements of galaxy
clustering, galaxy-galaxy lensing, and the galaxy SMF as a function of
redshift. Using this method we can derive $f_{\star}$ down to much
lower halo masses (a few $10^{11}$ M$_{\sun}$) than possible using
group and cluster catalogs. Second, we use a group membership catalog
to directly calculate $f_{\star}$ for $10^{13}<M_h/M_{\odot}<10^{14}$
by summing together the stellar masses of group
members. Interestingly, we find results that are significantly
different than previous work. Assuming a Chabrier Initial Mass
Function (IMF), our results suggest that previously published
estimates of $f_{\star}$ on group scales are overestimated by a factor
of about 2 to 5. The discrepancy is only partially reduced by assuming
a Salpeter IMF. We investigate the cause of this discrepancy and find
that previous work has overestimated $f_{\star}$ due to a combination
of inaccurate stellar mass estimators and/or because they have assumed
that all galaxies in groups and clusters are passive early type
galaxies with a constant mass-to-light ratio.

The layout of this paper is as follows. The data are briefly described
in Section \ref{cosmos_survey} followed by the presentation of our
method used to derive $f_{\star}$ in Section \ref{section_method}. Our
main results are presented in Section \ref{results}. Finally, we
discuss the results and draw up our conclusions in Section
\ref{conclusions}.

We assume a WMAP5 $\Lambda$CDM cosmology with $\Omega_{\rm m}=0.258$,
$\Omega_\Lambda=0.742$, $\Omega_{\rm b}h^2=0.02273$, $n_{\rm
  s}=0.963$, $\sigma_{8}=0.796$, $H_0=72$ km~s$^{-1}$~Mpc$^{-1}$
\citep[][]{Hinshaw:2009}. We assume a WMAP5 cosmology to maintain
consistency with the HOD analysis of \citet[][]{Leauthaud:2011a},
which is an key ingredient for this paper. All distances are expressed
in physical Mpc units.  Halo mass is denoted $M_h$, and we define
$M_{500c}\equiv M(<r_{500c})=500\rho_{\rm crit} \frac{4}{3}\pi
r_{500c}^3$ where $r_{500c}$ is the radius at which the mean interior
density is equal to 500 times the critical density ($\rho_{\rm
  crit}$). Stellar mass, denoted $M_{*}$, is derived using a Chabrier
Initial Mass Function (IMF) unless otherwise specified.  The
integrated (total) stellar content is denoted $f_{\star}$ and is
defined as $f_{\star}\equiv M_*^{\rm tot}/M_{500c}$.  Herein,
$f_{\star}$ is the sum of two components: a contribution from
satellite galaxies denoted $f_{\star}^{\rm sat}\equiv M_*^{\rm
  sat}/M_{500c}$ and a contribution from central galaxies denoted as
$f_{\star}^{\rm cen}\equiv M_*^{\rm cen}/M_{500c}$; it does not
include the ICL.  The function $E(z) \equiv H(z)/H_0 =
\sqrt{\Omega_{m}(1+z)^3+\Omega_{\Lambda}}$ represents the Hubble
parameter evolution for a flat metric. Stellar mass scales as
$1/H_0^2$. Halo mass scales as $1/H_0$. All magnitudes are given on
the AB system.


\section{Data description}\label{cosmos_survey}
 
The COSMOS survey \citep[][]{Scoville:2007}, centered at 10:00:28.6,
+02:12:21.0 (J2000), brings together a broad array of panchromatic
observations with imaging data from X-ray to radio wavelengths and a
large spectroscopic follow-up program (zCOSMOS) with the VLT
\citep{Lilly:2007}. In the following sections we describe the key
COSMOS data sets and previously published analyses relevant to this
paper.

\subsection{Stellar Mass Estimates}\label{stellar_masses}

The goal of this paper is to estimate the total stellar content of
dark matter halos. Stellar mass estimates are therefore an important
component of this paper. We adopt the same stellar masses as used by
\citet{Leauthaud:2011a} (hereafter ``L11'') and George et al. (2011)
(hereafter, ``Ge11''). These have been derived in a similar manner as
\citet{Bundy:2010} but are based on updated redshift information (v1.7
of the COSMOS photo-$z$ catalog and the latest available spectroscopic
redshifts as compiled by the COSMOS team) and use a slightly different
cosmology ($H_0=72$ km s$^{-1}$ Mpc$^{-1}$ instead of $H_0=70$ km
s$^{-1}$ Mpc$^{-1}$). We refer to L11 and \citet{Bundy:2010} for
details regarding the derivation of the stellar masses and only give a
brief description here. Stellar masses are calculated using COSMOS
ground-based photometry (filters $u^*, B_J, V_J, g^+, r^+, i^+, z^+,
K_s$) and the depth in all bands reaches at least 25th magnitude (AB)
with the $K_s$-band limited to $K_s < 24$. Stellar masses are
estimated using the Bayesian code described in \citet{Bundy:2006}
assuming a Chabrier Initial Mass Function (IMF) and the
\citet[][]{Charlot:2000} dust model. The stellar mass completeness
limits are shown in Figure 2 of L11 and vary from $10^{8.8}$
M$_{\odot}$ at $z=0.37$ to $10^{9.8}$ M$_{\odot}$ at $z=0.88$.

\subsection{Cosmos X-ray group membership catalog}\label{groups}

The entire COSMOS region has been mapped through 54 overlapping {\sl
  XMM-Newton} pointings and additional {\sl Chandra} observations
cover the central region (0.9 degrees$^2$) to higher resolution
\citep[][]{Hasinger:2007,Cappelluti:2009, Elvis:2009}. These X-ray
data have been used to construct a COSMOS X-ray group catalog that
contains 211 extended X-ray sources over 1.64 degrees$^2$, spanning
the range $0<z<1$, and with a rest-frame 0.1--2.4 keV luminosity range
between $10^{41}$ and $10^{44}$ erg s$^{-1}$. The general data
reduction process can be found in \citet{Finoguenov:2007} and details
regarding improvements and modifications to the original catalog are
given in \citet[][]{Leauthaud:2010}, George et al. (2011), and
Finoguenov in prep. Halo masses have been measured by
\citet[][]{Leauthaud:2010} for this sample by using weak lensing to
calibrate the relationship between X-ray luminosity (L$_X$) and halo
mass. Groups in this catalog have halo masses that span the range
$10^{13} \lesssim M_{h}/M_{\sun} \lesssim 10^{14}$.

To estimate $f_{\star}$ from this group catalog, we need a method to
distinguish galaxies that belong to groups from those in the
field. For this we use the group membership probability catalog from
Ge11. In Ge11, the full Probability Distribution Function (PDF) from
the COSMOS photo-$z$ catalog \citep[][]{Ilbert:2009} has been used to
derive a group membership probability for all galaxies with
$F814W<24.2$ from the COSMOS ACS galaxy catalog
\citep[][]{Leauthaud:2007,Leauthaud:2011a}. The completeness and
purity of the group membership assignment was studied in detail by
Ge11 using spectroscopic redshifts and mocks catalogs. Within a radius
of $0.5R_{200c}$, the group membership catalog has an estimated purity
of 84\% and a completeness of 92\%. The completeness and purity of
this catalog does not vary strongly with redshift at $z<1$.

\subsection{Cosmos analysis of galaxy-galaxy lensing, clustering, and
  stellar mass functions}\label{hod_analysis}

In L11 a HOD model was used to probe the relationship between halo
mass and galaxy stellar mass from $z=0.2$ to $z=1$. Constraints were
obtained by performing a joint fit (in three redshift bins) to
galaxy-galaxy lensing, galaxy clustering, and the galaxy stellar mass
function, with a 10 parameter HOD model. A detailed description of
this model can be found in \citet{Leauthaud:2011}. In particular, as
described in Section 2.3 of that paper, this model can be used to
derive $f_{\star}$ as a function of $M_h$. Hereafter, we refer to this
approach as the ``HOD method''. This method is discussed in more
detail in Section \ref{hod_method}.


\section{Method}\label{section_method}

The aim of this paper is to derive $f_{\star}$ as a function of
$M_{h}$. To achieve this goal, we use two different and complementary
measurements. First, we use a HOD model that has been calibrated to
fit measurements of galaxy clustering, galaxy-galaxy lensing, and the
galaxy SMF as a function of redshift. Second, we use a group
membership catalog to directly calculate $f_{\star}$ by summing
together the stellar masses of group members. There are advantages and
disadvantages inherent to each method. The HOD method might be
considered to be a more indirect probe of $f_{\star}$, however, it
allows us to derive $f_{\star}$ over a much wider halo mass range than
with group and cluster catalogs. For example, in COSMOS, this
technique allows us to probe $f_{\star}$ down to the stellar mass
completeness limit of the survey ($M_{*}\simeq 7\times10^{8}$
M$_{\odot}$ at $z=0.37$). The group catalog method is more direct and
less prone to modeling uncertainties (for example, the assumption that
satellite galaxies follow the same radial distributions as the dark
matter). However, this technique is limited by the halo mass range for
which group and cluster catalogs can be constructed ($M_h \gtrsim
10^{13}$ M$_{\odot}$) and is subject to other uncertainties such as
the accurate identification of the centers of groups and cluster of
galaxies. The group catalog method is also subject to uncertainties
due to the completeness and purity of the membership assignment and to
projection effects, that must be understood through mock catalogs.
Finally, we compare both methods with the results from the abundance
matching technique described in \cite{Behroozi:2010}, with the
assumption of the COSMOS stellar mass function.  This model has
substantially fewer parameters but makes a specific assumption about
the connection between galaxies and dark matter subhalos.

\subsection{HOD method}\label{hod_method}

The relationship between halo mass and galaxy stellar mass can be
described via a statistical model of the probability distribution
$P(N|M_h)$ that a halo of mass $M_h$ is host to N galaxies above some
threshold in luminosity or stellar-mass. This statistical model,
commonly known as the halo occupation distribution (HOD), has been
considerably successful at interpreting the clustering properties of
galaxies \citep[e.g.,][]{Seljak:2000, Peacock:2000, Scoccimarro:2001,
  Berlind:2002, Bullock:2002, Zehavi:2002, Zehavi:2005, Zheng:2005,
  Zheng:2007, Tinker:2007,Wake:2011,Zehavi:2011,White:2011}. Although
the HOD is usually inferred observationally by modeling measurements
of the two-point correlation function of galaxies, $\xi_{gg}(r)$, it
can also be used to model other observables such as galaxy-galaxy
lensing and the stellar-mass function. In L11 the HOD model was used
to probe the relationship between halo mass and galaxy stellar mass
from $z=0.2$ to $z=1$. Constraints were obtained by performing a joint
fit to galaxy-galaxy lensing, galaxy clustering, and the galaxy
stellar mass function.

In this paper, we adopt the parameter fits from Table 5 in L11 and we
use these to study $f_{\star}$ as a function of $M_h$. We calculate
$f_{\star}$ following the procedure outlined in Section 2.3 of
\citet{Leauthaud:2011}. One potential concern with this method is that
the calculation of $f_{\star}^{\rm sat}$ might require extrapolating
the HOD model beyond the lower and upper stellar mass bounds for which
the model has been calibrated. As shown in L11, $f_{\star}$ is
dominated by the stellar mass of central galaxies for $M_h \lesssim
10^{13}$ M$_{\odot}$. This concern is therefore only valid for halos
larger than about $10^{13}$ M$_{\odot}$ where satellite galaxies
represent the dominant contribution to $f_{\star}$.

We now investigate the stellar mass range of satellite galaxies that
build up the bulk of $f_{\star}^{\rm sat}$ (the satellite contribution
to $f_{\star}$) in halos above $M_h \gtrsim 10^{13}$ M$_{\odot}$. The
original HOD analysis of L11 used $M_{200b}$ halo masses (defined with
respect to 200 times the mean matter density $\bar{\rho}$). For
simplicity, for the tests in this paragraph we will also use
$M_{200b}$ but the conclusions are also valid for $M_{500c}$
masses. Let us consider the stellar fraction for satellites in a fixed
stellar mass bin, $f_{\star}^{\rm sat}(M_h|M_*^{t1},M_*^{t2})$, where
$M_*^{t1}$ represents a lower stellar mass limit and $M_*^{t2}$
represents an upper stellar mass limit. The expression for
$f_{\star}^{\rm sat}(M_h|M_*^{t1},M_*^{t2})$ is given by:

\begin{equation}
f_{\star}^{\rm sat}(M_h|M_*^{t1},M_*^{t2})=\frac{1}{M_h}\int_{M_*^{t1}}^{M_*^{t2}}\Phi_s(M_*|M_h)M_*{\rm d}M_* .
\label{tot_msat_1}
\end{equation}

\noindent where $\Phi_s$ represents the conditional stellar mass
function for satellite galaxies \citep[\eg,][]{Leauthaud:2011}. Using
this expression, we have tested how $f_{\star}^{\rm
  sat}(M_h|M_*^{t1},M_*^{t2})$ varies with the integral limits
$M_*^{t1}$ and $M_*^{t2}$. We find that at fixed halo mass, the bulk
of $f_{\star}^{\rm sat}$ comes from galaxies with $M_{*}>10^{9}$
M$_{\odot}$. For example, for $M_{200b}>10^{13}$ M$_{\odot}$, 95$\%$
and 90$\%$ of $f_{\star}^{\rm sat}$ is due to satellites with $M_{*}
\gtrsim 2\times10^{9}$ M$_{\odot}$ and $M_{*} \gtrsim 7\times10^{9}$
M$_{\odot}$ respectively. The HOD model of L11 is calibrated down to
$6\times 10^8$ M$_{\odot}$ at $z=0.37$ and to $6\times 10^9$
M$_{\odot}$ at $z=0.88$. Therefore, the bulk of $f_{\star}^{\rm sat}$
arises from satellites that are well within the tested limits of our
model (for a similar calculation, also see \citealt{Puchwein:2010}).

As mentioned previously, the L11 analysis assumes $M_{200b}$ halo
masses whereas in this paper we use $M_{500c}$ halo masses so as to
facilitate comparisons with previous work. We therefore need to
convert the L11 results from $M_{200b}$ to $M_{500c}$. For this we
make two assumptions. First, we assume that satellites galaxies are
distributed according to the same profiles as their parent halos,
namely, NFW profiles \citep{Navarro:1997} with $C_{\rm sat}=C_{\rm
  halo}$ \citep[][]{Nagai:2005,Wetzel:2010,Tinker:2011}. Second, we
assume that the distribution of satellite galaxies is independent of
stellar mass (i.e, there is no ``satellite mass segregation''). This
assumption is supported by multiple observations
\citep[][]{Pracy:2005,Hudson:2010,von-der-Linden:2010,Wetzel:2011} but
see also \citet[][]{van-den-Bosch:2008}. Under these two assumptions,
$f_{\star}^{\rm sat}$ is unchanged by any halo mass conversion
(because the same conversion factor applies to both the numerator and
the denominator in $f_{\star}^{\rm sat}$). The $f_{\star}^{\rm cen}$
component does vary, however, because the halo mass (denominator)
changes while the mass of the central galaxy (numerator) remains
constant. For details on how to convert halo masses assuming a NFW
profile, see Appendix C in \citet{Hu:2003a}. For this conversion, we
assume the mass-concentration relation of
\citet[][]{Munoz-Cuartas:2011}.

\subsection{Group catalog method}\label{group_cat_method}

\begin{figure}[thb]
\epsscale{1.15}
\plotone{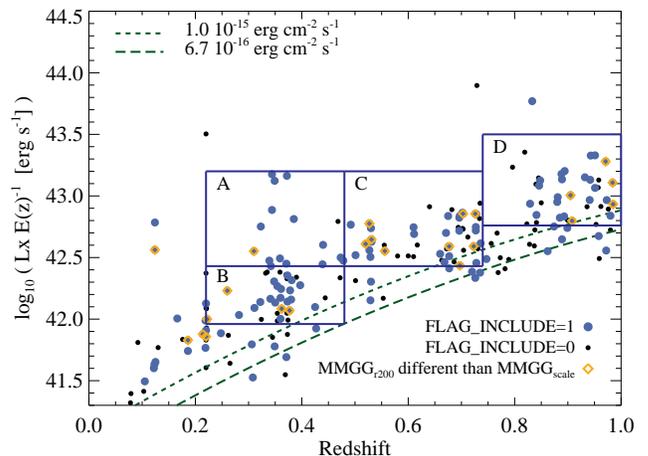}
\caption{Cosmos X-ray group sample as a function of redshift and
  $L_XE(z)^{-1}$, where $L_X$ represents X-ray luminosity and $E(z)
  \equiv H(z)/H_0$. Large blue circles indicate groups included in
  this analysis, and small black circles show groups that are excluded
  from this analysis (see text for selection flags). Yellow diamonds
  show systems where the centering is deemed uncertain based on the
  analysis of George et al. in prep., because the most massive group
  member lies far from the X-ray center. Blue boxes indicate the binning
  scheme. The redshift limits for this binning scheme are chosen to
  match the HOD analysis of L11.}
\label{xgroup_bin}
\end{figure}

In this section, we describe our procedure for calculating $f_{\star}$
at group scales using the COSMOS X-ray group catalog. Figure
\ref{xgroup_bin} shows the group sample as a function of redshift and
$L_X$. Blue boxes show the binning scheme that we adopt for this
paper. The redshift limits are selected to match those of L11 and are
$z_1=[0.22,0.48]$, $z_2=[0.48,0.74]$, and $z_3=[0.74,1]$.

To ensure a robust group sample with a relatively clean member list,
we include only groups with reliable optical associations having
greater than 3 members, and we exclude close neighboring systems and
groups near the edges of the field or masked regions where member
assignment is difficult (catalog flags \textsc{xflag}=1 or 2, and
\textsc{poor}=\textsc{merger}=\textsc{mask}=0; see Ge11 for flag
definitions). This ensemble of quality cuts is synthesized by a global
flag in Ge11 called \textsc{flag\_include} (we select groups with
\textsc{flag\_include}=1). In total, after these cuts, our catalog
contains 129 groups at $z<1$.
 
The determination of group centers is an important and non-trivial
task, especially since these structures do not always have a visually
obvious central galaxy. Mis-centering effects could lead to an
underestimate of $f_{\star}$, especially if the mis-centering is large
enough that the true central galaxy is excluded from the group
selection for example. George et al. in prep show that weak lensing is
an excellent tool that can be used to optimize centering algorithms by
maximizing the stacked weak lensing signal at small radii (below about
500 kpc). For example, George et al. in prep also show that various
naive centering algorithms such as the centroid of galaxy members,
even if weighted by stellar mass or luminosity, is a poor tracer of
the centers of dark matter halos.  We use the group centers from
George et al. in prep that are found to maximize the stacked weak
lensing signals at small radii, namely, the most massive group galaxy
located within an NFW scale radius of the X-ray centroid, denoted
MMGG$_{\rm scale}$.  We note, however, that George et al. in prep have
also identified a sub-set of groups (of order 20\%) that host a more
massive galaxy at a greater distance from the X-ray centroid than the
NFW scale radius. For these groups, the weak lensing does not clearly
indicate a preferred center among the two options. We include these
systems in our analysis but we have tested that our results are not
affected by this choice.

Group members are selected according to the following criteria:

\begin{enumerate}
\item $P_{\rm mem}>0.5$ where $P_{\rm mem}$ denotes the group
  membership probability derived in Ge11.
\item $r_{2d}<R_{500c}$ where $r_{2d}$ is the projected distance to
  the group center (defined hereafter by the position of the
  MMGG$_{\rm scale}$).
\item Galaxies must have a stellar mass above the completeness limit
  of the bin in consideration (see Table \ref{bin_schem}).  Stellar
  mass completeness limits for the group membership catalog are higher
  than those quoted in Section \ref{stellar_masses}, because the group
  membership catalog is limited to $F814W<24.2$.
\end{enumerate}

Group members are selected only within a radius of $R_{500c}$ for two
reasons. Ge11 have shown that the purity of the group membership
assignment drops at larger radii (see their Figure 5). By only
considering galaxies at $r_{2d}<R_{500c}$, we reduce the errors due to
purity corrections.  In addition, most previously published work on this
topic has used $r_{2d}<R_{500c}$; this choice thus facilitates
comparisons with previous work (see \S \ref{comparison}).

Once group members have been selected following the criteria outlined
above, $f_{\star}$ is derived for each group as the sum of the stellar
masses of all group members associated with this group divided by the
halo mass. Halo masses are derived from the X-ray luminosity following
the weak lensing scaling relation of \citet[][]{Leauthaud:2010}.

Although in principle more direct than the HOD method, measuring
$f_{\star}$ estimates from the group catalog has systematic errors
that must be corrected for. The two main effects on the measurement
are:

\begin{enumerate}
\item The completeness and purity of the membership selection. Here we
  include both contamination in the group membership selection due to
  neighboring galaxies but also the de-projection of a spherical NFW
  profile. Indeed, the quantity that we are interested in is
  $f_{\star}$ contained within a sphere of radius $R_{500c}$ whereas
  our membership selection is akin to selecting galaxies in a cylinder
  in redshift space. We therefore include a correction factor for this
  de-projection.
\item The stellar mass completeness of the group membership catalog
  (i.e., we need to account for the contribution to $f_{\star}$ from
  galaxies below the completeness limit).
\end{enumerate}

To derive these correction factors, we use a suite of COSMOS-like mock
catalogs that have been described in Ge11. Mocks were created from a
single simulation (named ``Consuelo'') 420 $h^{-1}$ Mpc on a
side\footnotemark[1]\footnotetext[1]{In this paragraph, numbers are
  quoted for $H_0=100$ h km~s$^{-1}$~Mpc$^{-1}$. The assumed cosmology
  for Consuelo is $\Omega_{\rm m}=0.25$, $\Omega_\Lambda=0.75$,
  $\Omega_{\rm b}h^2=0.02273$, $n_{\rm s}=1.0$, $\sigma_{8}=0.8$,
  $H_0=70$ km~s$^{-1}$~Mpc$^{-1}$.}, resolved with 1400$^3$ particles,
and a particle mass of 1.87$\times 10^{9}$ $h^{-1}$ M$_{\odot}$. This
simulation can robustly resolve halos with masses above $\sim 10^{11}$
$h^{-1}$ M$_{\odot}$ which corresponds to a central galaxy stellar
masses of $\sim 10^{8.5}$ $h^{-1}$ M$_{\odot}$, well-matched to our
completeness limit of $F814W=24.2$ at low redshift. This simulation is
part of the Las Damas suite\footnotemark[2]\footnotetext[2]{Details
  regarding this simulation can be found at {\tt
    http://lss.phy.vanderbilt.edu/lasdamas/simulations.html}} (McBride
et al. in prep). Halos are identified within the simulation using a
friends-of-friends finder with a linking length of $b=0.2$, and
populated with mock galaxies according to the HOD model of L11. See
Ge11 for further details regarding these mock catalogs.

We derive a single overall correction factor to $f_{\star}$ due to the
two effects described above. To do so, we apply the group membership
selection to the mock catalog and we compare the measured value of
$f_{\star}$ (in the mocks) to the true value of $f_{\star}$ (within a
sphere of radius $R_{500c}$). We derive correction factors separately
for the contribution to $f_{\star}$ from satellite and central
galaxies. These correction factors are denoted $C_{\rm sat}$ and
$C_{\rm cen}$ respectively. $C_{\rm sat}$ is defined as:

\begin{equation}
C_{\rm sat} = \log_{10}(M_*^{\rm mock\_truth}/M_*^{\rm mock\_sel}) 
\end{equation}

\noindent where $M_*^{\rm mock\_truth}$ represents the ``true'' total
mass from the mocks in a sphere of $R_{500c}$ and down to the
completeness limit of the mocks and $M_*^{\rm mock\_sel}$ represents
our group member selection applied to the mocks. $C_{\rm cen}$
represents the equivalent for central galaxies. The correction factors
are listed in Table \ref{corr_fac} for the three redshift bins and are
at most 14\%. The completeness and purity correction component of
$C_{\rm sat}$ is negative (the measured value of $f_{\star}^{\rm sat}$
is higher than it should be because of contamination) whereas the
stellar mass completeness component of $C_{\rm sat}$ is positive (the
measured value of $f_{\star}^{\rm sat}$ must be augmented to account
for satellites below our completeness limit). The magnitude of the two
corrections is fairly similar, but the opposite signs mean that these
corrections tend to cancel each other out. The redshift trends cause
the net correction for $C_{\rm sat}$ to be negative at $z\sim
0.37$ and positive at $z\sim 0.88$.

\begin{deluxetable*}{lllll}{htb}
\tablecolumns{5} \tablecaption{Group binning scheme\label{bin_schem}} \tablewidth{0pt} 
\startdata
\hline 
\hline 
\\  [-1.5ex]
Bin ID & Redshift & lower L$_X$ limit & upper L$_X$ limit & M$_*$ completeness limit\tablenotemark{a} \\ 
    &          & $\log_{10}($L$_X$E(z)$^{-1}$/erg s$^{-1}$)  &$\log_{10}($L$_X$E(z)$^{-1}$/erg s$^{-1}$) & $\log_{10}$(M$_*$/M$_{\odot}$) \\ [1ex]
\hline\\[-1.5ex]
A & $z_1=[0.22,0.48]$ & 41.96 & 42.43 & 9.3  \\ 
B & $z_1=[0.22,0.48]$ & 42.43 & 43.2 & 9.3  \\ 
C & $z_2=[0.48,0.74]$ & 42.43 & 43.2 & 9.9  \\ 
D & $z_3=[0.74,1]$ & 42.76  & 43.5 & 10.3  \\ 
\enddata
\tablenotetext{a}{The stellar mass completeness limits cited here are
  for the group membership catalog.}
\end{deluxetable*}

\begin{deluxetable}{lllll}  
\tablecolumns{4} \tablecaption{Correction factors for group catalog \label{corr_fac}} \tablewidth{0pt} 
\startdata
\hline 
\hline 
\\  [-1.5ex]
Bin ID & Redshift & $C_{\rm sat}$ & $C_{\rm cen}$ & \\ 
\hline\\[-1.5ex]
A & $z_1=[0.22,0.48]$ & -0.029 & -0.019   \\ 
B & $z_1=[0.22,0.48]$ & -0.066 & -0.004  \\ 
C & $z_2=[0.48,0.74]$ & -0.047  &  -0.017  \\ 
D & $z_3=[0.74,1]$ & 0.003  &  -0.017  \\ 
\enddata
\end{deluxetable}

Due to the finite mass resolution of our mock catalogs, the
contribution to $f_{\star}$ from very low mass galaxies (those that
are not contained in our mock catalogs) will not be included. Using
Equation \ref{tot_msat_1}, we estimate that this contribution should
only be of order 1-2$\%$, and we thus neglect it here.


\section{Results}\label{results}

\subsection{Results from the HOD method}\label{results_hod}

\begin{figure*}[thb]
\epsscale{0.8}
\plotone{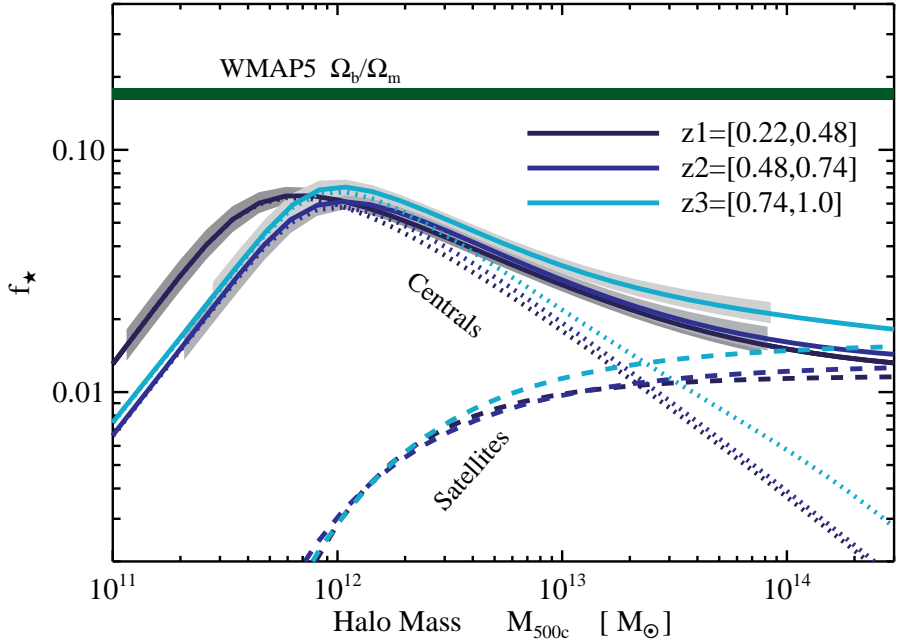}
\caption{Total stellar content in galaxies (not including ICL) as a
  function of halo mass.  Results from the HOD method constrained by
  COSMOS data are compared to the cosmic baryon fraction measured by
  the Wilkinson Microwave Probe \citep[WMAP5;
  $f_b=\frac{\Omega_b}{\Omega_m}=0.171\pm 0.009$;][]{Dunkley:2009}.
  The results for three redshift ranges $z_1$, $z_2$, and $z_3$ are
  shown by the solid dark blue, blue, and turquoise lines
  respectively. The shaded region shows statistical errors
  only. Systematic errors are quantified in Section \ref{sys_error}
  (also see Figure \ref{sys_err}).  $f_{\star}$ is dominated by
  central galaxies at $M_{500c} \lesssim 10^{13.2}$ M$_{\odot}$
  (dotted lines) and by satellites at $M_{500c} \gtrsim 10^{13.2}$
  M$_{\odot}$ (dashed lines).  }
\label{tot_sm_content}
\end{figure*}

We now calculate $f_{\star}$ using the best fit HOD parameters from
L11 for each of the three redshift bins. The results are shown in
Figure \ref{tot_sm_content}. The solid dark blue, blue, and turquoise
lines show $f_{\star}$ as a function of $M_{500c}$ for $z\sim 0.37$,
$z\sim 0.66$, and $z\sim 0.88$ respectively. The contribution to
$f_{\star}$ from central galaxies, $f_{\star}^{\rm cen}$, is shown by
the dotted red line and the contribution from satellites,
$f_{\star}^{\rm sat}$, is shown by the dashed yellow line. $f_{\star}$
is dominated by central galaxies at $M_{500c} \lesssim 10^{13.2}$
M$_{\odot}$ and by satellites at $M_{500c} \gtrsim 10^{13.2}$
M$_{\odot}$.  Note that $f_{\star}$ in this figure does not include
ICL.

The redshift evolution of $f_{\star}$ has been discussed at length in
L11. Briefly, the ``pivot halo mass'' is defined as the halo mass
where $f_{\star}$ reaches a maximum ($M_{500c} \simeq 10^{12}$
M$_{\odot}$). At fixed halo mass and below the pivot halo mass scale,
$f_{\star}$ increases at later epochs. This evolution can be explained
by the fact that in this regime, central galaxy growth outpaces halo
growth. At fixed halo mass and above the pivot halo mass scale, there
is tentative evidence that $f_{\star}$ declines at later epochs. In
L11, we suggest that this trend might be linked to the smooth
accretion of dark matter, which brings no new stellar mass, and
amounts to as much as 40\% of the growth of dark matter halos
\citep[e.g.,][]{Fakhouri:2010} and/or the destruction of satellites
and a growing ICL component.

The shaded grey area in Figure \ref{tot_sm_content} indicates the
statistical error on $f_{\star}$. The statistical error is small
compared to the systematic uncertainty associated with the actual
derivation of stellar masses. We will return to this issue shortly in
Section \ref{sys_error}.

\subsection{Results from the group catalog}\label{results_group}

\begin{figure*}[thb]
\epsscale{1.15}
\plotone{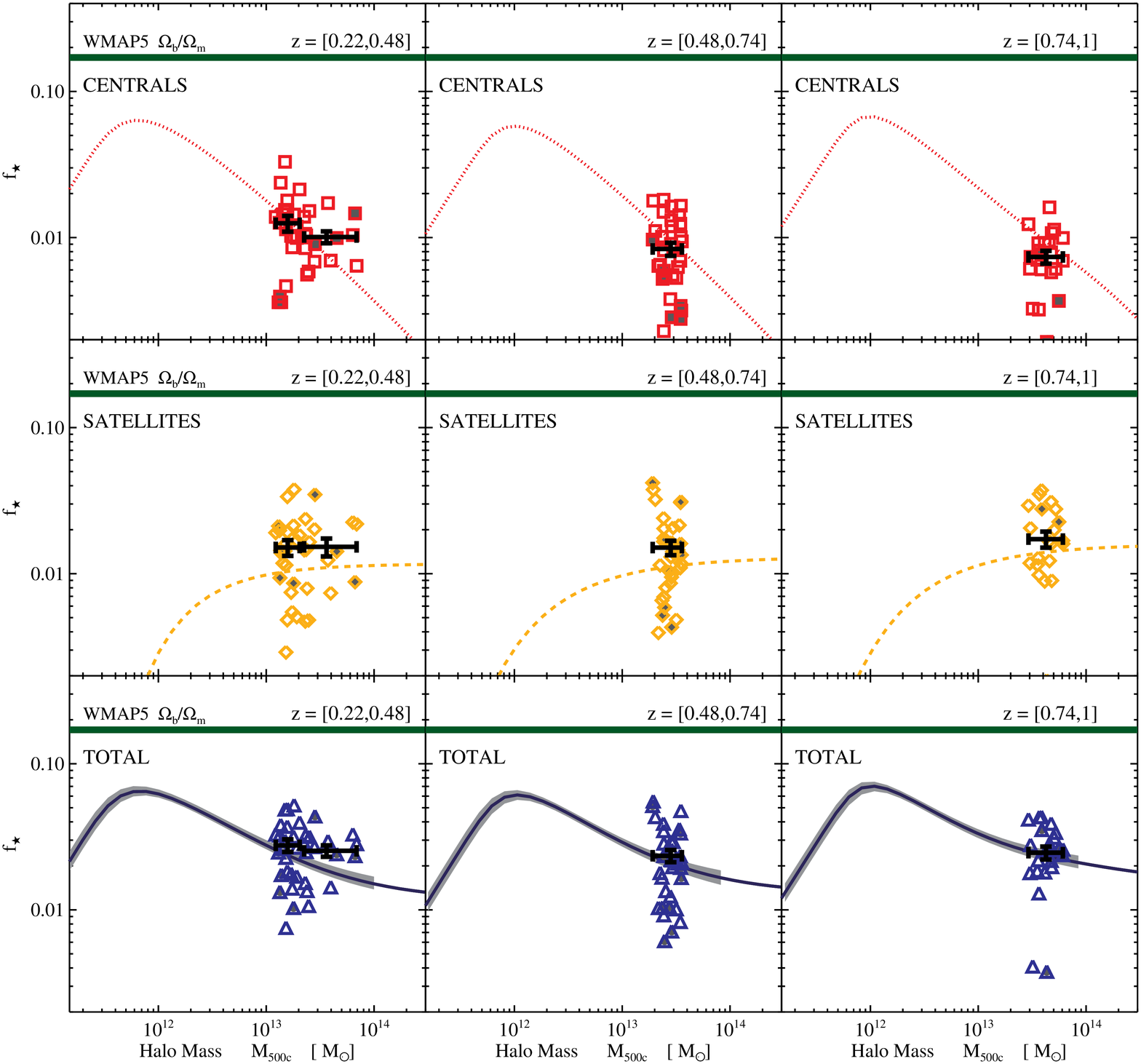}
\caption{Comparison between $f_{\star}$ inferred from our HOD model
  with results from the COSMOS X-ray group catalog as a function of
  redshift.  The HOD values for $f_{\star}^{\rm cen}$, $f_{\star}^{\rm
    sat}$, and $f_{\star}$ are shown respectively by the dotted red
  line, and the dashed yellow line, and solid blue line.  The red
  squares in the three top panels show $f_{\star}^{\rm cen}$
  measurements from the COSMOS X-ray group catalog (each data point
  represents one X-ray group) for the three redshift bins. Black data
  points with errors show the mean value in each bin. The horizontal
  error corresponds to the width of the bin while the vertical error
  represents $\sigma/\sqrt{N_g}$.  Here $\sigma$ is the measured
  dispersion in the data and $N_g$ is the number of groups in the
  bin. Yellow diamonds in the three middle panels show $f_{\star}^{\rm
    sat}$,  and blue diamonds in the lower three panels show
  $f_{\star}$=$f_{\star}^{\rm cen}$+$f_{\star}^{\rm sat}$. Grey solid
  points indicate systems for which the centering is ambiguous
  (excluding these systems does not change our results). }
\label{tot_sm_content_500c_matt}
\end{figure*}

We now calculate $f_{\star}$ at group scales using the COSMOS X-ray
group catalog. Figure \ref{tot_sm_content_500c_matt} shows our
estimates for $f_{\star}^{\rm cen}$ (top row), $f_{\star}^{\rm sat}$
(middle row), and $f_{\star}$ (bottom row) as derived from the group
catalog (the correction factors from Table \ref{corr_fac} have been
applied). Each colored data point in Figure
\ref{tot_sm_content_500c_matt} represents a measurement from one X-ray
group. The thick black data point represents the mean value in each
halo mass bin (see Table \ref{bin_schem}). The horizontal error
corresponds to the width of the bin while the vertical error
represents $\sigma/\sqrt{N_g}$ where $\sigma$ is the measured
dispersion in the bin and $N_g$ is the number of groups in the bin.

The solid filled symbols in Figure \ref{tot_sm_content_500c_matt}
indicate systems for which the weak lensing analysis of George et
al. in prep suggests that the group centering might be unreliable
(MMGG$_{\rm scale}$ is different than MMGG$_{\rm r200}$). We do not
find a strong correlation between these systems and either halo mass
or $f_{\star}$, and find that the results are unchanged if these
systems are removed from the analysis.

We find excellent agreement between the HOD method and the group
catalog method for $f_{\star}^{\rm cen}$ (compare the red squares to
the red dotted line). Regarding $f_{\star}^{\rm sat}$, there is a
small but consistent offset between the two methods (compare the
yellow diamonds to the yellow dashed line). Indeed, in all three
redshift bins, the $f_{\star}^{\rm sat}$ values as measured from the
group catalog are higher than those derived from the HOD method by
about 20\% to 50\%. As mentioned in the previous section, the HOD
predictions for $f_{\star}^{\rm sat}$ in Figure
\ref{tot_sm_content_500c_matt} have been derived under the assumption
that the distribution of satellite galaxies follows the same NFW as
their parent halos. We hypothesize that the offset for $f_{\star}^{\rm
  sat}$ seen in Figure \ref{tot_sm_content_500c_matt} might indicate
that either a) the satellite mass distribution has different
concentration than the dark matter or b) the satellite mass
distribution varies with stellar mass (``mass segregation''). The
observed level of offset, however, does not affect the main
conclusions of this paper and so we leave a more detailed
investigation of this effect for future work.

Overall, the two methods show a remarkable level of agreement
especially considering the fact that they are subject to very
different types of systematic errors. The two methods agree to within
30\% for the estimated value for $f_{\star}$ at group scales.


\subsection{Systematic errors associated with $f_\star$
  measurements}\label{sys_error}

The statistical uncertainty on $f_{\star}$ (from both the HOD method
or the group catalog) is much smaller than the systematic uncertainty
associated with the determination of stellar masses. For example,
according to \citet[][]{Behroozi:2010}, the typical systematic error
associated with photometric-based stellar mass estimates are of order
$0.25$ dex. This systematic uncertainty arises from the choice of a
stellar population synthesis (SPS) model, the choice of a dust
attenuation model, and the assumed functional form of the star
formation history. In addition, there are also further uncertainties
due to the choice of an IMF. Further details regarding systematic
errors in stellar mass measurements can be found in
\citet{Behroozi:2010} and \citet{Conroy:2009a}.

In this section, we attempt to quantify how the systematic errors
described above might translate into an uncertainty on
$f_{\star}$. Note that the aim here is only to provide a first attempt
to discuss systematic errors on $f_{\star}$ due to uncertainties in
stellar mass estimates (to our knowledge, this has not yet been
quantified). We will make certain assumptions that will undoubtedly
need to be refined in future work. We focus on the low redshift
results here ($z\lesssim 0.4$) because the method used here
relies on the availability of a large number of published SMFs.

To begin with, we note that the SMF provides the strongest
observational constraints on the HOD model of L11. The other two
observables used by L11 (clustering and galaxy-galaxy lensing) contain
additional information, but this is less dominant than the information
from the SMF. In this paper, we will therefore make the simplifying
assumption that the main source of systematic error in the
determination of $f_{\star}$ is due to uncertainties related to the
SMF. Under this assumption, the task of determining systematic errors
on $f_\star$ now becomes one of determining systematic errors on the
observed SMF. We consider two distinct sources of error for the SMF:

\begin{enumerate}
\item A systematic error due to the choice of an IMF. In this paper,
  we specifically consider the Salpeter and the Chabrier IMF
  \citep{Salpeter:1955,Chabrier:2003}. To first order, the choice of a
  Chabrier versus a Salpeter IMF, yields masses that are lower by
  about 0.25 dex\footnotemark[3]\footnotetext[3]{In practice, this
    conversion depends on the adopted SPS model and also the star
    formation history. The conversion can vary at the 0.05 dex level:
    $\log_{10}(M_*^{Sal}/M_*^{Chab})=0.25 \pm0.05$ dex.}. For
  simplicity, we will assume here that we can simply convert between
  these two IMFs using a difference of 0.25 dex and we will refer to
  this difference as the ``IMF systematic error on the SMF''. Note
  that another IMF that is often considered is the Kroupa IMF
  \citep{Kroupa:2001}. The Kroupa IMF yields stellar masses that are
  larger than a Chabrier IMF by about 0.05 dex. We neglect any
  possible variations of the IMF with either galaxy type and/or
  redshift but note that this is another potential (and currently
  poorly determined) source of systematic error on $f_{\star}$.
\item All other sources of systematic error on the SMF, for example,
  those due to the choice of a SPS model, the choice of a dust attenuation model,
  and the assumed functional form of the star formation history. We
  will refer to these together as the ``non-IMF systematic errors on
  the SMF''.
\end{enumerate}

\begin{figure}[thb]
\epsscale{1.2}
\plotone{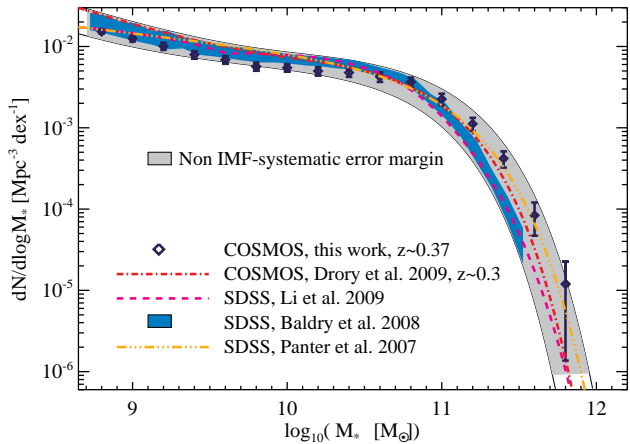}
\caption{Compilation of stellar mass functions at $z<0.4$. Systematic
  error on the SMF not including the IMF is indicated by the shaded
  grey region.  Variation due to measurement error, sample variance
  between surveys, and redshift evolution is implicitly included, but
  we do not explore the full range of SPS models and SPS model
  parameters.}
\label{smf_comp}
\end{figure}

To estimate the non-IMF systematic errors, we consider a compilation
of various published SMFs at $z<0.4$. Figure \ref{smf_comp} shows an
ensemble of low-z ($z<0.4$) mass functions from COSMOS
\citep[][]{Drory:2009,Leauthaud:2011a} and from the SDSS
\citep[][]{Panter:2007,Baldry:2008,Li:2009}. All mass functions in
Figure \ref{smf_comp} have been converted to our assumed value of
$h=0.72$ and to a Chabrier IMF. Observed variations between these
different mass functions are due to a combination of systematic error,
measurement error\footnotemark[4]\footnotetext[4]{A higher level of
  measurement error will lead to an inflated stellar mass function at
  the high mass end due to Eddington bias; see for example discussion
  in \citealt{Behroozi:2010}.}, sample variance, and redshift
evolution\footnotemark[5]\footnotetext[5]{Most studies find very
  little redshift evolution in the total SMF out to $z=1$
  \citep[\eg,][]{Bundy:2006}. Therefore, although there is a
  considerable time span between $z=0$ and $z=0.4$, the differences
  between COSMOS and SDSS SMFs over this time period are likely
  primarily due to systematic errors and not redshift evolution.}. We will
adopt a conservative approach and assume that all the observed
variation in Figure \ref{smf_comp} is due to non-IMF systematic errors
in the determination of stellar masses (this is an assumption that
will need to be improved on in future work). We define a lower
envelope and an upper envelope that is designed to encompass the
observed range of SMFs. This is shown by the grey shaded region in
Figure \ref{smf_comp}. In what follows, we assume that this grey
shaded region represents the non-IMF systematic errors on the
SMF. There is, however, one caveat with this method. Our non-IMF
systematic error margin might be underestimated due to the fact that
not all SPS models are represented by the set of SMFs shown in Figure
\ref{smf_comp}\footnotemark[6]\footnotetext[6]{\citet[][]{Panter:2007}
  use \citet[][]{Bruzual:2003} SPS models. \citet{Baldry:2008} use the
  average of four stellar mass estimates from
  \citet[]{Kauffmann:2003a}, \citet[]{Panter:2004},
  \citet[]{Glazebrook:2004}, and
  \citet[]{Gallazzi:2005}. \citet{Li:2009} use masses from
  \citet{Blanton:2007}. \citet{Drory:2009} and this paper use
  \citet[][]{Bruzual:2003} SPS models.}. For example, the SPS models
of \citet[][]{Maraston:2005} are not represented in Figure
\ref{smf_comp}. To date, however, there are no published $z<0.4$ SMFs
that have used the \citet[][]{Maraston:2005} models. We therefore
simply note that this is a limitation with our current method that
needs to be improved on in future work.

\begin{figure}[thb]
  \epsscale{1.15}
\plotone{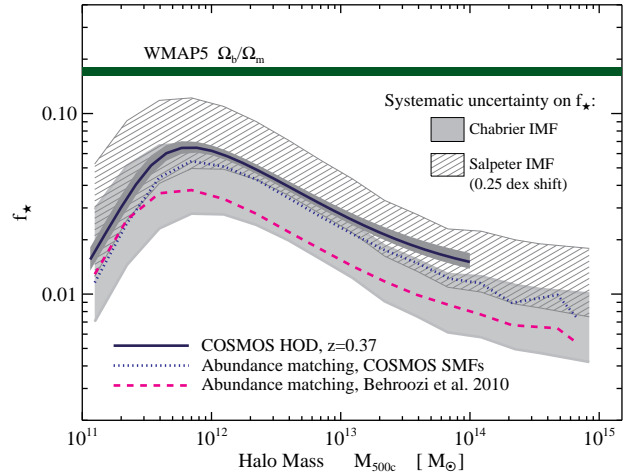}
  \caption{Systematic error on $f_{\star}$ due to uncertainties in
    stellar mass estimates. The dark blue line shows the COSMOS HOD
    result at $z\sim0.37$. The dotted blue line shows $f_{\star}$
    derived by applying the same abundance matching technique as
    \citet[][]{Behroozi:2010} to the COSMOS $z\sim0.37$ SMF. The HOD
    method and the abundance matching method yield slightly difference
    results at the 15\% to 20\% level. The dashed magenta line shows
    $f_{\star}$ as measured by \citet[][]{Behroozi:2010} by abundance
    matching to the \citet[][]{Li:2009} SDSS SMF. The difference
    between the COSMOS HOD result (dark blue solid line) and the
    \citet[][]{Behroozi:2010} result (dashed magenta line) is largely
    driven by differences between the COSMOS mass function and the
    \citet[][]{Li:2009} mass function (see Figure \ref{smf_comp}). The
    solid light grey region shows our the systematic error margin for
    $f_{\star}$, assuming a Chabrier IMF. The fact that the COSMOS
    results lie above this region at $M_{500c}>10^{13}$ $M_{\odot}$ is
    due to a $15\%$ model uncertainty between the HOD method and the
    abundance matching method. The hashed region shows the systematic
    error margin for $f_{\star}$ assuming a Salpeter IMF (note that
    the hashed region simply corresponds to a 0.25 dex upward shift of
    the solid light grey region).}
\label{sys_err}
\end{figure}

The next step is to understand how these non-IMF errors on the SMF
might translate into a systematic error on $f_{\star}$. At present,
the HOD model of L11 can only be used in conjunction with clustering
and galaxy-galaxy lensing measurement. For this reason, we do not use
the L11 HOD method for this step. Instead, we will adopt an
``abundance matching'' methodology that assumes there is a monotonic
correspondence between halo mass (or circular velocity) and galaxy
stellar mass (or luminosity)
\citep[e.g.,][]{Kravtsov:2004,Vale:2004,Tasitsiomi:2004,Vale:2006a,Conroy:2006,Conroy:2009,Moster:2010,Behroozi:2010,Guo:2010}. Abundance
matching techniques are generally designed to fit the SMF alone and
have fewer free parameters than the HOD method. We follow the
abundance matching techniques of \citet[][]{Behroozi:2010}, using
$M_{500c}$ halo masses, using both the upper and the lower limits of
the grey shaded region in Figure \ref{smf_comp}.  The abundance
matching results are then reported in Figure \ref{sys_err} as the
non-IMF systematic error margin for $f_{\star}$ (solid light grey
region).  The systematic errors margin for $f_{\star}$ assuming a
Salpeter IMF are estimated by shifting this region upwards by 0.25 dex
in stellar mass (hashed region in Figure \ref{sys_err}).

In Figure \ref{sys_err}, the COSMOS results are slightly higher than
our non-IMF systematic error margin in Figure \ref{sys_err} at
$M_{500c}>10^{13}$ M$_{\odot}$.  This is due to the fact that the HOD
method of L11 and the abundance matching method of
\citet[][]{Behroozi:2010} yield slightly different predictions for
$f_{\star}$ as function of halo mass; this is discussed further in the
following section.

\subsection{Comparison with the abundance matching results of
  \citet{Behroozi:2010}}\label{comparison_beh}

\citet{Behroozi:2010} have derived constraints on $f_{\star}$ (see
their Figure 10) by abundance matching to the SDSS SMF of
\citet[][]{Li:2009}.  Figure \ref{sys_err} shows these results, here
translated to $M_{500c}$ halo masses (magenta dashed line) and
assuming a Chabrier IMF.  The \citet{Behroozi:2010} results yield a
lower amplitude for $f_{\star}$ than the L11 COSMOS results. This is
largely driven by differences between the COSMOS mass function and the
\citet[][]{Li:2009} mass function (shown in Figure \ref{smf_comp}).

In order to investigate how much of these differences are in fact due
to differences in the assumed stellar mass function, we have applied
the abundance matching technique of \citet{Behroozi:2010} to the
COSMOS $z\sim0.37$ SMF.  This is shown as the blue dotted line in
Figure \ref{sys_err}.  This can then be directly compared with the L11
HOD results.  We find that the HOD results in roughly $15$\% higher
values for all halo masses than the abundance matching technique shown
here.

There are several possible sources for the 15\% discrepancy between
the HOD and abundance matching methods.  One possibility is that the
weak lensing and/or clustering is providing more information in the
HOD analyses than is available given the stellar mass function alone,
and that this information pushes the values slightly higher.  Another
potential issue is that conversions have been made in each case to use
$M_{500c}$ halo masses (HOD results were calculated initially using
$M_{200b}$ and abundance matching was done using $M_{vir}$); these
conversions necessarily make assumptions about the radial profiles of
satellites which may be inaccurate at the several percent level.
Another possibility is that the abundance matching method might suffer
from satellite incompleteness in the N-body simulations (see e.g.  Wu
et al. in prep), however, this is unlikely to be important below the
group mass scale.  We leave tracking down the exact source of the
difference to future work, and simply note that there is roughly a
$15$\% level uncertainty in the determination of the non-IMF
systematic uncertainty on $f_{\star}$ due to the method used to derive
$f_{\star}$ from the SMF.  Overall, the difference between L11 and
\citet{Behroozi:2010} is consistent with our systematic error margin
for $f_{\star}$, and these differences do not impact any of our
conclusions significantly.

\subsection{Comparison with previous results from group and cluster
  catalogs}\label{comparison}

\begin{figure*}[htb]
\epsscale{1.19}
\plotone{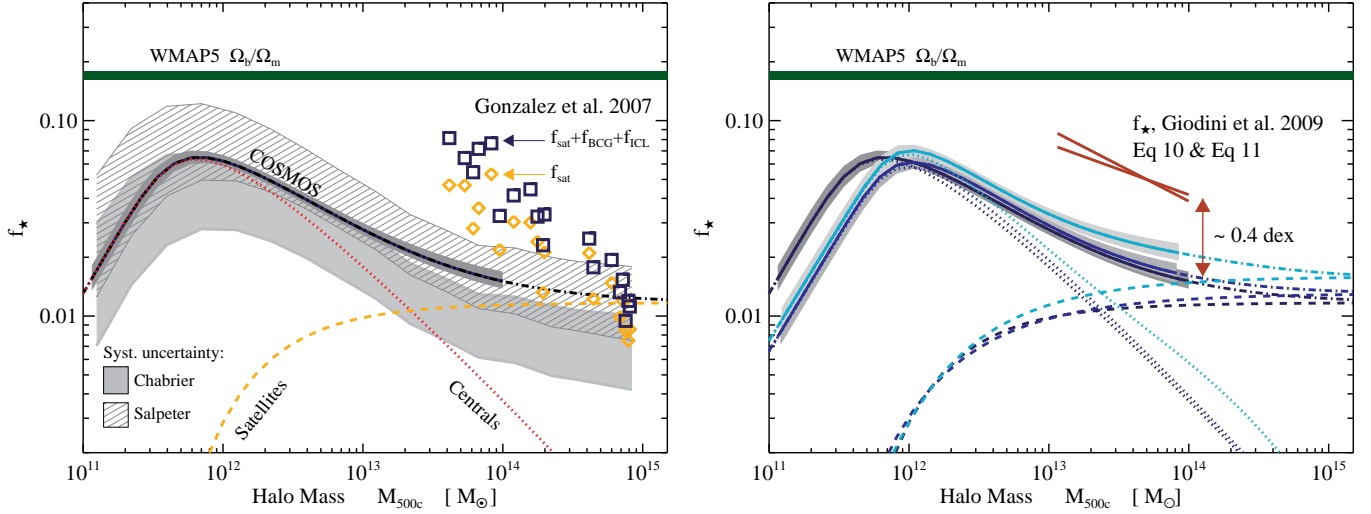}
\caption{Same as Figure \ref{tot_sm_content}, now including
  $f_{\star}$ measurements from Gz07 (left panel) and Gi09 (right
  panel). The solid light grey region in the left panel shows our the
  systematic error margin for $f_{\star}$, assuming a Chabrier IMF
  (see $\S$ \ref{sys_error}). The hashed region shows the systematic
  error margin for $f_{\star}$ assuming a Salpeter IMF. Yellow
  diamonds in the left hand panel represent data-points from Gz07 for
  the satellite contribution to $f_{\star}$. This component can be
  directly compared to our low redshift satellite term (yellow
  dash-dash line). The total stellar mass fraction from Gz07 is shown
  by the blue squares ($f_{\star}^{\rm sat}$+$f_{\star}^{\rm
    BCG}$+$f_{\star}^{\rm ICL}$). Note that our predictions do not
  include the ICL contribution: our blue solid line corresponds to
  $f_{\star}^{\rm sat}$+$f_{\star}^{\rm BCG}$. The right panel shows
  the total galaxy stellar mass fraction ($f_{\star}^{\rm
    sat}$+$f_{\star}^{\rm BCG}$, red solid line) measured by Gi09. The
  Gi09 fit is averaged over all groups at $z<1$ and so is 
  most comparable to our $z_2$ bin (solid blue line). In both cases,
  our results suggest significantly lower values for $f_{\star}$ than
  reported by either of these previous studies.}
\label{tot_sm_content_500c}
\end{figure*}

We now compare our $f_{\star}$ measurements with two recently
published results from \citet{Gonzalez:2007} (hereafter ``Gz07'') and
\citet{Giodini:2009} (hereafter ``Gi09'') which have been derived
using group and cluster catalogs. Figure \ref{tot_sm_content_500c}
shows our predictions for $f_{\star}$ as a function of $M_{500c}$
compared to the measurements of Gz07 and Gi09. In both cases, our
results suggest {\em significantly lower values} for $f_{\star}$ than
reported by either of these previous studies. We now provide a more
detailed comparison with both Gz07 and Gi09 and discuss possible
explanations for the source of the discrepancy.

\subsubsection{Comparison with \citet{Gonzalez:2007}}\label{comparison_gonzalez}

The left panel of Figure \ref{tot_sm_content_500c} shows the
comparison of our low-z results with Gz07. Yellow diamonds show
estimates from Gz07 for the contribution to $f_{\star}$ from satellite
galaxies\footnotemark[7]\footnotetext[7]{$f_{\star}^{\rm sat}$ is
  calculated from Table 1 (column 7) in \citet{Gonzalez:2007} using a
  Vega solar luminosity of $M_{\rm sun}=3.94$ (A.Gonzalez, priv. comm)
  and a Cousins I-band mass-to-light ratio of $M/L_I=3.6$.} whereas
dark blue squares show the total stellar fraction ($f_{\star}^{\rm
  sat}$+$f_{\star}^{\rm BCG}$+$f_{\star}^{\rm ICL}$). The yellow
diamonds in Figure \ref{tot_sm_content_500c} are therefore directly
comparable to our prediction for $f_{\star}^{\rm sat}$ which is shown
by the yellow dash-dash line. Gz07 find that $f_{\star}^{\rm sat}$
increases towards lower halo masses whereas we predict the opposite
trend. It is clear from this figure that a large fraction of the
observed discrepancy is due to the $f_{\star}^{\rm sat}$ component
(and not so much to $f_{\star}^{\rm BCG}$ or $f_{\star}^{\rm
  ICL}$). We will now investigate the source of this difference in
further detail.

To estimate stellar masses, Gz07 have used a Cousins I-band
mass-to-light ratio ($M/L_I=3.6$) calibrated from the dynamical
modeling of 2-d kinematic data from \citet{Cappellari:2006}. The
\citet{Cappellari:2006} sample is comprised of early type galaxies
from the SAURON sample (Bacon et al. 2001). We note two important
caveats with this approach:

\begin{enumerate}
\item The SAURON galaxy sample is comprised solely of elliptical (E)
  and lenticular (S0) galaxies which have larger $M/L$ values on
  average than intermediate and late type galaxies. If the early type
  fraction increases with halo mass as suggested by several studies
  \citep[\eg,][]{Weinmann:2006, Hansen:2009, Wetzel:2011}, then this
  could explain both why Gz07 find a steeper slope and a higher
  amplitude than we do for $f_{\star}^{\rm sat}$ at group scales.
\item The $M/L$ ratios presented in \citet{Cappellari:2006} are {\em
    dynamical} (i.e total) and are therefore sensitive to the dark
  matter fraction within the effective radius ($R_e$). The dynamical
  mass-to-light ratios in \citet{Cappellari:2006} might therefore be
  biased high by up to $30\%$ compared to the stellar mass-to-light
  ratios (see Figure 17 in \citealt{Cappellari:2006} for example).
\end{enumerate}

\begin{figure*}[htb]
\epsscale{1.1}
\plotone{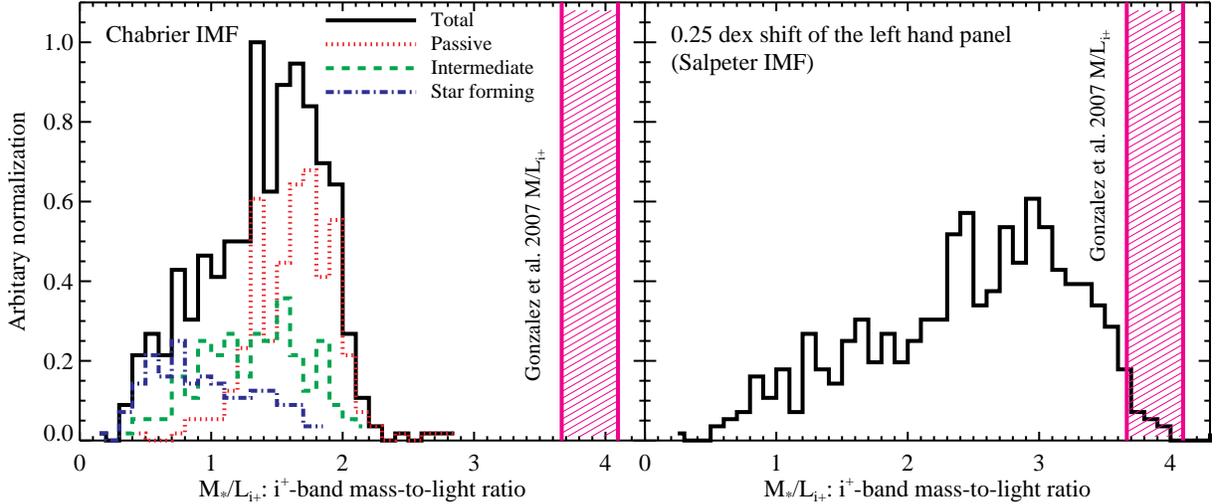}
\caption{{\em Left}: $i^+$-band mass-to-light ratios of low-z group
  members as a function of galaxy color as inferred from our stellar
  mass estimates (Chabrier IMF). The mass-to-light ratio used by Gz07
  is shown by the hashed magenta region (there is a spread in values
  due to to conversion from Cousins I to Subaru $i^+$). Assuming a
  Chabrier IMF, the stellar masses inferred by using the Gz07
  $M/L_{i^+}$ are larger than ours by a factor of 2 to 10. The
  difference is particularly large for active (blue dash-dot line) and
  intermediate type (green dashed line) galaxies. {\em Right}:
  $i^+$-band mass-to-light ratios assuming stellar masses are
  increased by 0.25 dex. To first order, these are the mass-to-light
  ratios expected for a Salpeter IMF. The difference is reduced by
  assuming a Salpeter IMF but the Gz07 M/L ratio is still well above
  the mean M/L ratio of our group galaxy sample.}
\label{ml_comp}
\end{figure*}

We now investigate the mass-to-light ratios of our group members
compared to the value used by Gz07. For this exercise, we consider
group members from the membership catalog of Ge11 in our two low
redshift bins (bins A and B in Figure \ref{xgroup_bin}). We derive
Subaru $i^+$-band mass-to-light ratios from our stellar mass catalog
using the absolute $i^+$-band AB magnitude that is provided by the
COSMOS photoz catalog. For this exercise we adopt an AB solar
luminosity of $M_{\rm sun}=4.54$\footnotemark[8]\footnotetext[8]{This
  value is adopted from
  http://mips.as.arizona.edu/~cnaw/sun.html}. Figure \ref{ml_comp}
shows the $M/L_{i^+}$ values for COSMOS group members. We use
unextincted rest frame magnitudes from the COSMOS photoz catalog to
divide the histogram in Figure \ref{ml_comp} by galaxy color. More
specifically, we define galaxy color as $C=M(NUV)-M(R)$ where $M(NUV)$
and $M(R)$ are the unextincted rest-frame template magnitudes in the
near-ultraviolet and the R bands defined in
\citet[][]{Ilbert:2010}. We adopt the same division as in that paper,
namely:

\begin{center}
$C<1.2$ ``high activity'' or ``blue''

$1.2<C<3.5$ ``intermediate activity'' or ``intermediate''

$C>3.5$ ``quiescent'' or ``red''
\end{center}

We compute the difference in magnitudes between our Subaru $i^+$
filter and a Cousins I filter for a range of stellar population
templates. We find that a Cousins $M/L_I=3.6$ corresponds to a Subaru
$i^+$ mass-to-light ratio of $3.67<M/L_{i+}<4.1$ (the exact value
depends on galaxy color). These values are represented in Figure
\ref{ml_comp} by the hashed magenta region.

Figure \ref{ml_comp} shows that the stellar masses inferred by using a
similar $M/L$ as Gz07 are larger than our SED stellar masses by a
factor of 2 to 10 (assuming a Chabrier IMF). The difference is
particularly large for active and intermediate type galaxies. The
large spread observed in Figure \ref{ml_comp} highlights the fact that
the group population is not well represented by any single $i^+$-band
$M/L_{i+}$ ratio value. Part of the observed spread can also be
explained by the sensitivity of the $i^+$-band luminosity to recent
star formation, underscoring the need for NIR data to more accurately
determine galaxy stellar mass. The differences between our
mass-to-light ratios and the value used by Gz07 is reduced by assuming
a Salpeter IMF (Figure \ref{ml_comp}, right panel). Even with a
Salpeter IMF, however, the Gz07 M/L ratio is still well above the mean
M/L ratio of our galaxy sample. This is likely to be due to the fact
that the \citet{Cappellari:2006} mass-to-light ratios are dynamical
and are thus sensitive to dark matter in addition to stellar mass.

Another important fact worth highlighting in Figure \ref{ml_comp} is
that a non negligible fraction of group members are star
forming/intermediate type galaxies (also see Ge11 who find that the
quenched fractions in groups is of order 40\% to 60\% at
$z<0.5$). Note however that Figure \ref{ml_comp} does not include the
corrections described in Section \ref{group_cat_method} for
de-projection or the completeness and purity of the membership
selection. These corrections were derived as adjustments to the total
$f_{\star}$ from mock catalogs that do not distinguish quiescent
galaxies from active ones, so we cannot directly apply them to the
separate galaxy types in Figure \ref{ml_comp}. To address the
possibility of color-dependent contamination of the group sample, we
can test the purity and completeness of the membership selection using
a subsample of galaxies with spectroscopic redshifts as done in
Ge11. The COSMOS spectroscopic redshift sample, however, is dominated
by zCOSMOS ``bright'' galaxies with $F814W<22.5$ and is therefore not
representative of our group membership sample ($F814W<24.2$). This
will induce a degree of uncertainty that can only be reduced with
deeper and more representative spectroscopic data and/or more
representative mock catalogs. Having noted this caveat, we use this
approach to compute the completeness and purity of the membership
selection within $R_{500c}$ and as a function of galaxy type. We find
that the group populations of red, green, and blue galaxies are
overestimated by 14\%, 20\%, and 30\% respectively. After applying
these correction factors to groups in bins A and B (and for galaxies
above our stellar mass completeness limit), we find that the mean red
fraction at $z<0.48$ is roughly 50\%. So while we do see some added
contamination of our group member sample among blue galaxies, they
still represent a significant part of the population. Studies that aim
to compute $f_{\star}$ must take this varied group population into
account. It is clear that assuming that all group members are
quiescent could lead to large biases in $f_{\star}$ estimates.

\begin{figure}[thb]
\epsscale{1.25}
\plotone{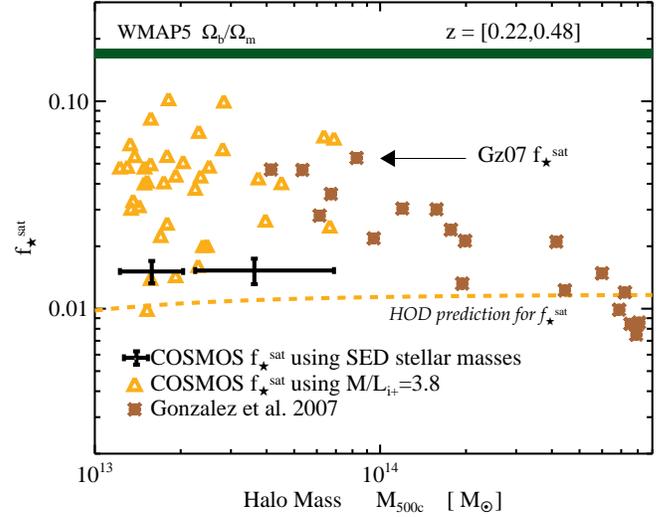}
\caption{Impact of mass-to-light ratio assumptions on the fraction of
  stars in satellites, $f_{\star}^{\rm sat}$.  In addition to the
  default values from our group catalog (black error bars), values
  obtained from our group catalog using a similar $i^+$-band
  mass-to-light ratio as Gz07 ($M/L_{i^+}=3.8$) are shown (yellow
  triangles).  In detail the two samples are not directly comparable
  because our sample has lower halo masses and is at higher redshifts
  than Gz07.  However, the fact that the Gz07 results are
  qualitatively reproduced using $M/L_{i^+}=3.8$ suggests that the
  derivation of stellar masses is the primary cause of the observed
  difference between our results and Gz07.}
\label{mlratiotest}
\end{figure}

Finally, we now investigate the impact of using a single mass-to-light
value similar to the one use by Gz07. For this exercice we adopt a
single value of $M/L_{i^+}=3.8$ and we neglect the small color
dependance of the $M/L$ conversion from Cousins I to Subaru $i^+$ (see
Figure \ref{ml_comp}). Using the COSMOS group catalog, we re-compute
$f_{\star}^{\rm sat}$ using the absolute $i^+$-band AB magnitude
provided by the COSMOS photoz catalog and a mass-to-light ratio of
$M/L_{i^+}=3.8$ to derive stellar masses. Here, we only consider
$f_{\star}^{\rm sat}$. Because our analysis does not include ICL, we
cannot compare with the Gz07 $f_{\star}^{\rm BCG}$+$f_{\star}^{\rm
  ICL}$ component (these were measured together as a single component
by Gz07). However, as noted in Figure \ref{tot_sm_content_500c}, most
of the discrepancy between our results and Gz07 occurs for
$f_{\star}^{\rm sat}$. Figure \ref{mlratiotest} shows that we are able
to qualitatively reproduce the Gz07 $f_{\star}^{\rm sat}$ using our
data and $M/L_{i^+}=3.8$. Note however that the halo mass range and
redshift range of our sample does not exactly match Gz07 and so Figure
\ref{mlratiotest} is not exactly a one-to-one comparison. However, the
fact that we obtain similar $f_{\star}^{\rm sat}$ values as Gz07
strongly suggests that the primary cause of the discrepancy is due the
fact that $M/L_{i^+}=3.8$ largely overestimates the stellar masses of
group members compared to our full SED fitting technique.

\subsubsection{Comparison with \citet{Giodini:2009}}\label{comparison_giodini}

Gi09 have used the COSMOS data and an earlier version of our X-ray
catalog to derive $f_{\star}$. Since we use the same data set and a
similar group catalog, our results are directly comparable. The right
panel of Figure \ref{tot_sm_content_500c} shows the comparison of our
results with Gi09. Since Gi09 report the average over all groups at
$z<1$, their results are most comparable with our $z_2$ redshift
bin. To estimate stellar masses, Gi09 have used a galaxy-type
dependent stellar mass-to-$K_s$ band luminosity ratio derived by using
the analytical relation from \citet[][]{Arnouts:2007} assuming a
Salpeter IMF. However, as pointed out by \citet{Ilbert:2010}, the
\citet[][]{Arnouts:2007} $M/L_{K_s}$ relation was only calibrated for
massive galaxies. In practice the K-band mass-to-light ratio varies
with galaxy age and color. Thus the \citet[][]{Arnouts:2007}
$M/L_{K_s}$ relation overestimates the stellar masses of low mass
galaxies. \citet{Ilbert:2010} have shown that the
\citet[][]{Arnouts:2007} mass estimates contain galaxy-color dependent
biases compared to SED fitting techniques (see their Figure 28). For
example, \citet{Ilbert:2010} find that the \citet[][]{Arnouts:2007}
masses are biased high by more than 0.3 dex for star forming galaxies
with $\log_{10}(M_*/M_{\odot})<9.5$. Thus, including these galaxies
will lead to an overestimate of $f_{\star}$.

\begin{figure}[thb]
\epsscale{1.25}
\plotone{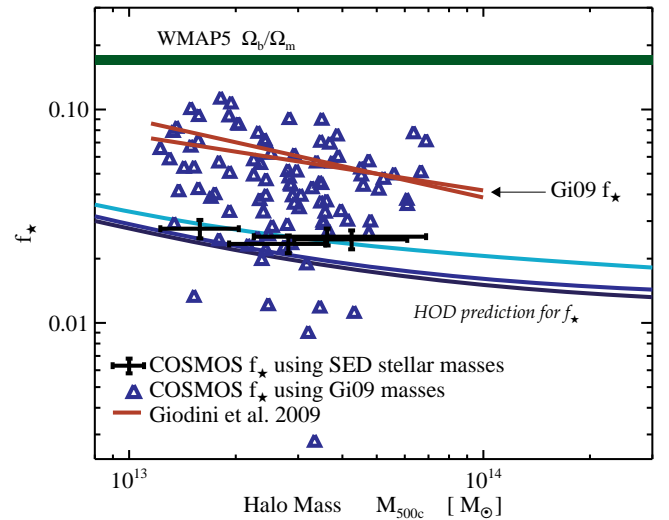}
\caption{Impact of stellar mass estimates on $f_{\star}$. In addition
  to the default values from our group catalog (black error bars),
  values obtained from our group catalog using the same
  \citet[][]{Arnouts:2007} stellar mass estimates as Gi09 are shown
  (blue triangles).  These are in qualitative agreement with the Gi09
  results (red lines), indicating that the assumptions about stellar
  masses are the primary source of the difference between the two sets 
 of results. }
\label{mlratiotest2}
\end{figure}

Since the Gi09 results are based on COSMOS data, we can make a direct
comparison between their mass estimates and ours. We match our galaxy
catalog with the photoz catalog that was used by Gi09. We then
recompute $f_{\star}$ from our catalog using the same procedure as in
Section \ref{results_group} but using the \citet[][]{Arnouts:2007}
masses and the results are shown in Section \ref{results_group}. For
this exercise we have used the same redshift and halo mass bins as in
Section \ref{results_group} (see Table \ref{bin_schem}). Note however
that we do not apply the purity/completeness and de-projection
correction factors here. Indeed, since the Gi09 stellar masses are
different than ours, we would need a new set of mock catalogs to
perform this correction. The correction values however are typically
only of order 5 to 14\% and are much smaller than difference between
our results and Gi09 (a factor of $\sim$2.5). Neglecting these
correction factors here should therefore not change our main
conclusions.

Figure \ref{mlratiotest2} shows that we obtain roughly similar values
for $f_{\star}$ as Gi09 when we use the \citet[][]{Arnouts:2007}
masses. In detail, however, we do not exactly find the same mean
values as a function of halo mass as Gi09 but our methods are
sufficiently different (Ge11 use a Bayesian probabilistic method
whereas Gi09 use a background subtraction method) that this is not too
unexpected. Part of the observed difference in Figure
\ref{mlratiotest2} is due to the fact that Gi09 use a Salpeter IMF
whereas we use a Chabrier IMF (this accounts for about 0.25 dex). The
remaining difference (about 0.15 dex) is probably due to a combination
of differences in the techniques that we have used and to differences
in stellar mass estimates.


\section{Discussion and Conclusions}\label{conclusions}

In this paper we have derived the total stellar mass fraction,
$f_{\star}$, as a function of host halo mass from $z=0.2$ to $z=1$
using HOD methods, abundance matching methods, and direct estimates
from group catalogs. Our stellar masses are derived from full SED
fitting to multi-band photometry (including K-band) from the COSMOS
survey. Assuming a Chabrier IMF, we find significantly lower estimates
for $f_{\star}$ at group scales than previous work. Including
(non-IMF) systematic errors on stellar masses, our analysis suggests
that $0.012\lesssim f_{\star} \lesssim 0.025$ at $M_{500c}=10^{13}$
M$_{\odot}$ and $0.0057\lesssim f_{\star} \lesssim 0.015$ at
$M_{500c}=10^{14}$ M$_{\odot}$. We will make files available upon
request including our $f_{\star}$ estimates and the associated
systematic uncertainties. Our main results are as follows:

\begin{enumerate}
\item Assuming a Chabrier IMF, we find that previously published
  estimates of $f_{\star}$ on group scales could be overestimated by a
  factor of two to five. This discrepancy is only partially reduced by
  assuming a Salpeter IMF. We investigate the cause of this
  discrepancy and find that a large fraction of the observed
  difference can be explained by the use of over-simplistic
  mass-to-light ratio estimates.
\item We show that galaxies in groups are a mixed population of
  quiescent, intermediate, and star forming galaxies. As a
  consequence, galaxies in groups are not well represented by any
  single $M/L$ ratio value. The assumption that all galaxies in groups
  are quiescent, for example, will lead to biases in $f_{\star}$
  estimates.
\item We quantify the systematic uncertainty on $f_{\star}$ using
  abundance matching methods and show that the statistical
  uncertainty on $f_{\star}$ ($\sim 10$\%) is currently dwarfed by
  systematic uncertainties associated with stellar mass measurements
  ($\sim 45$\% excluding IMF uncertainties). We provide first
  estimates for the systematic error margin on $f_{\star}$ due to
  uncertainties in stellar mass measurements. While we have tried to
  provide a conservative estimate for these systematic errors, it is
  also clear that this is a non-trivial task that will require further
  investigation. One aspect in particular that could be improved on in
  this respect would be to consider a larger set of SMFs that cover a
  more representative range of SPS models than we have used here.
\item While direct measurements of $f_{\star}$ using group and cluster
  catalogs are necessary and important, HOD and abundance matching
  methods can probe $f_{\star}$ over a much wider halo mass range than
  possible using group and cluster catalogs. For example, in COSMOS,
  HOD methods allows us to probe $f_{\star}$ down to the stellar mass
  completeness limit of the survey ($M_* \sim7\times10^{8}$
  M$_{\odot}$ and $M_h\sim 10^{11}$ M$_{\odot}$ at $z=0.37$). In
  COSMOS, we show that the HOD method and the group catalog method are
  in good overall agreement but exhibit some small but interesting
  differences regarding the contribution to $f_{\star}$ from satellite
  galaxies ($f_{\star}^{\rm sat}$).
\item Using identical data sets, we find that the HOD method of
  \citet[][]{Leauthaud:2011a} and the abundance matching method of
  \citet[][]{Behroozi:2010} agree to within 15\% in terms of
  determining $f_{\star}$. Identifying the cause of remaining small
  differences between the two methods will require further
  investigation, but these small differences do not affect the
  main conclusions of this paper.
\end{enumerate}

Our finding that the total stellar fraction in groups is lower than
previously estimated has interesting implications for the
thermodynamic history of the intra-group gas. In simulations, the
relative proportions of gas mass to stellar mass are directly related
to the efficiency of cooling, star formation, and AGN feedback
\citep[\eg,][]{Kravtsov:2005, Puchwein:2008, Bode:2009,
  Puchwein:2010}. Suppressed gas mass fractions at small radii might
indicate that gas has been removed from the centers of groups and
clusters by non gravitational processes. For example, recent
simulations from \citet{McCarthy:2011} have suggested that low entropy
gas might be ejected from the progenitors of present-day groups by AGN
feedback at high redshift (primarily $2<z<4$) at the epoch when
super-massive black holes are in quasar mode. On the other hand, low
gas mass fractions might also be exactly compensated for by
$f_{\star}$+$f_{\rm ICL}$ in such a way that the total baryon fraction
is constant with halo mass and with halo radius. Discriminating
between these two scenarios is key towards understanding how much gas
may have been ejected from these systems.

\begin{figure*}[thb]
\epsscale{0.7}
\plotone{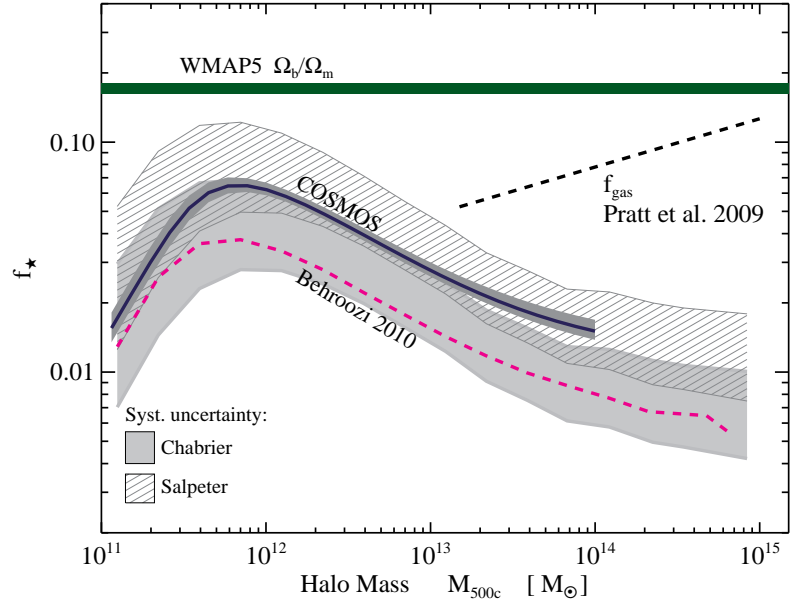}
\caption{Comparison of our $f_{\star}$ estimates with the mean gas
  mass fraction from \citet[][]{Pratt:2009}. Even assuming a Salpeter
  IMF, our measurements imply that the total stellar content of dark
  matter halos is always lower than the gas mass fraction for halos
  above $10^{13} M_{\odot}$ and for $r<R_{500c}$. }
\label{compare_sun}
\end{figure*}

\begin{figure*}[thb]
\epsscale{0.7}
\plotone{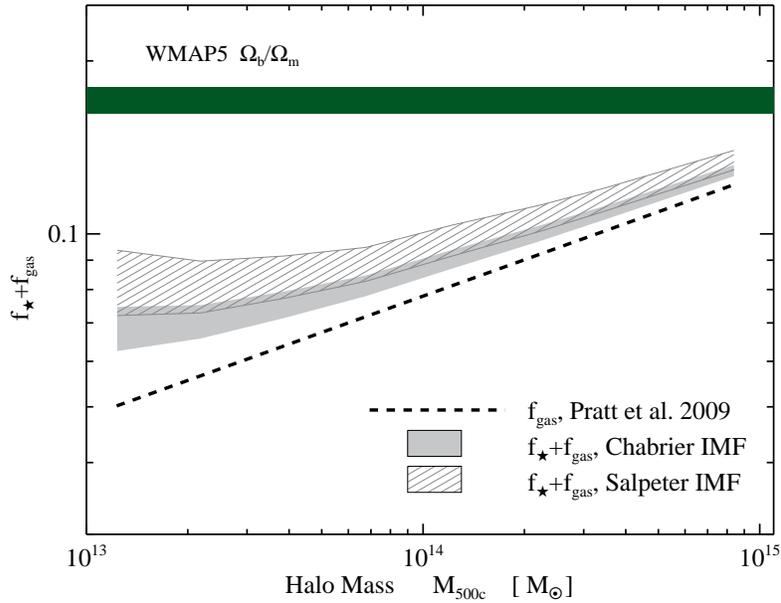}
\caption{Sum of the stellar fraction $f_{\star}$ and gas fraction
  $f_{\rm gas}$, where $f_{\rm gas}$ is the mean gas mass fraction
  taken from \citet[][]{Pratt:2009}. The solid grey region shows the
  systematic error margin for $f_{\star}$+$f_{\rm gas}$ assuming a
  Chabrier IMF. The grey hashed region shows the systematic error
  margin for $f_{\star}$+$f_{\rm gas}$ assuming a Salpeter IMF.}
\label{tot_f}
\end{figure*}

In Figure \ref{compare_sun}, we have compared our results with the
mean gas mass fraction ($f_{\rm gas}$) from \citet[][]{Pratt:2009},
computed from a compilation of data from \citet{Arnaud:2007},
\citet{Vikhlinin:2006}, and \citet[][]{Sun:2009}. This result
demonstrates that $f_{\star}$ is always lower than $f_{\rm gas}$ for
halos above $10^{13} M_{\odot}$, for the assumption of either a
Salpeter or Chabrier IMF.  Figure \ref{tot_f} shows the sum of
$f_{\star}$ and $f_{\rm gas}$.  Here the solid grey region indicates
the systematic error margin for $f_{\star}$+$f_{\rm gas}$ assuming a
Chabrier IMF; the grey hashed region shows the systematic error margin
for $f_{\star}$+$f_{\rm gas}$ assuming a Salpeter IMF.  Even in the
latter case, we find a shortfall in $f_{\star}$+$f_{\rm gas}$ compared
to the cosmic mean, which increases towards lower halos masses.

It is important to note that Figure \ref{tot_f} does {\em not} include
a baryonic contribution from ICL.  While including ICL in the baryon
budget is obviously an important issue to be adressed, $f_{\rm ICL}$
estimates suffer from large uncertainties related to the choice of a
$M/L$ value. For example, the ICL estimates of
\citet[][]{Gonzalez:2007} assume $M/L_I=3.6$; Figure \ref{ml_comp}
demonstrates this is on the high end compared to $M/L_I$ values
predicted from SPS models. Finally, we note that the main claim in
this paper, that $f_{\star}$ estimates are lower than previously
estimated, is dominated by the $f_{\star}^{\rm sat}$ component of
$f_{\star}$.

In summary, we have presented new measurements of the stellar fraction
$f_{\star}$ in groups and clusters from several complementary
approaches, which we believe to be more robust than previous estimates.
However, precise measurements of $f_{\star}$ are limited by our
systematic uncertainties associated with stellar mass
measurements. Future improvements to the $f_{\star}$ measurements
presented in this paper will benefit most from efforts that lead to an
improved understanding of galaxy stellar masses.


\acknowledgments

We thank Ian McCarthy, Andrey Kravtsov, Andrew Wetzel, and Stefania
Giodini for insightful discussions. We are grateful to Anthony
Gonzalez for useful comments on an early version of the manuscript and
for providing data in electronic format. We thank Michele Cappellari
and Claudia Maraston for providing details on IMF conversions and SPS
models. We are grateful to Jack Bishop and Al Leon for inspiring
conversations. AL acknowledges support from the Chamberlain Fellowship
at LBNL and from the Berkeley Center for Cosmological Physics. The HST
COSMOS Treasury program was supported through NASA grant
HST-GO-09822. We wish to thank Tony Roman, Denise Taylor, and David
Soderblom for their assistance in planning and scheduling of the
extensive COSMOS observations.  RHW and PSB received support from a
NASA HST Theory Grant HST-AR-12159.01-A and from the U.S. Department
of Energy under contract number DE-AC02-76SF00515, and used computing
resources at SLAC National Accelerator Laboratory.  We gratefully
acknowledge the contributions of the entire COSMOS collaboration
consisting of more than 70 scientists.  More information on the COSMOS
survey is available at {\bf
  \url{http://cosmos.astro.caltech.edu/}}. It is a pleasure the
acknowledge the excellent services provided by the NASA IPAC/IRSA
staff (Anastasia Laity, Anastasia Alexov, Bruce Berriman and John
Good) in providing online archive and server capabilities for the
COSMOS data-sets.
 
 
 


\bibliographystyle{apj}


\end{document}